\documentclass[epj]{svjour}
\usepackage{bm,graphicx,braket}
\usepackage{amsmath,amssymb}
\usepackage{upgreek}
\newcommand*\ii{\mathrm{i}}

\usepackage{cite}
\begin{document}
\title{{Topological edge states in the Su-Schrieffer-Heeger model subject to
  balanced particle gain and loss}}
\author{Marcel Klett\inst{1,2}, Holger Cartarius\inst{3}, Dennis Dast\inst{1},
  J\"org Main\inst{1}, \and G\"unter Wunner\inst{1}}%
%
%
\institute{
Institut f\"ur Theoretische Physik 1, Universit\"at Stuttgart,
70550 Stuttgart, Germany \and 
Institut f\"ur Theoretische Physik and Center for Quantum Science, 
Universit\"at T\"ubingen, Auf der Morgenstelle 14, 72076 T\"ubingen, Germany
\and Physik und ihre Didaktik, Universit\"at Stuttgart,
  70550 Stuttgart, Germany}
\date{Received: date / Revised version: date}
%
\abstract{
We investigate the Su-Schrieffer-Heeger model in presence of an injection
  and removal of particles, introduced via a master equation in Lindblad form.
  It is shown that the dynamics of the density matrix follows the predictions
  of calculations in which the gain and loss are modeled by complex
  $\mathcal{PT}$-symmetric potentials. In particular it is found that there
  is a clear distinction in the dynamics between the topologically
  different cases known from the stationary eigenstates.
\PACS{
    {03.65.Vf}{Phases: geometric; dynamic or topological}   \and
      {11.30.Er}{Charge conjugation, parity, time reversal, and other
        discrete symmetries}  \and
      {73.20.-r}{Electron states at surfaces and interfaces} \and
      {73.43.Nq}{Quantum phase transitions}
     } 
} 
\titlerunning{Topological edge states in the SSH model with balanced gain and loss}
\authorrunning{Marcel Klett et al.}
\maketitle
\section{Introduction}
\label{intro}

Today topological many-body systems are a strongly investigated topic
and are in many cases well understood \cite{Hasan2010a,Alicea12a,%
  Altland1997a,Schnyder2008a,Leijnse2012a}. One of the best known examples is
the explanation of the quantized Hall effect \cite{Klitzing1980a,Klitzing1986a}
in terms of a topological invariant \cite{Thouless1982a}. Of special interest
are topologically protected Majorana zero modes
\cite{Mourik2012a,Leijnse2012a,Stanescu2013a,Elliott2015a,Carmele2015a} since
they are robust against local defects or disorder. Typically two topologically
different phases can arise when a certain parameter of the system is varied.
In the topologically nontrivial phase (TNP) energies within the band gap
appear, and their corresponding eigenstates are called edge states since they
appear at the edge between two different types of solids or potentials. In the
topologically trivial phase (TTP) the gap closing states are absent.

Since no system is completely isolated recently the question was raised of how
a coupling to the environment can influence the existence of topologically
nontrivial modes and the appearance of topologically protected edge states
\cite{Hu2011a,Esaki2011a,Ghosh2012a,Viyuela2012a,Schomerus2013a,Rivas2013a,%
  Zeuner2015a,Yuce2015a,Zhu14a,Yuce2015b,Wang15,Yuce2016a,Benito2016a,%
  Niklas2016a,Sedlmayr2018a,Sedlmayr2018b}. Furthermore, it was shown that
dissipation can even be used to create topologically nontrivial states
\cite{Bardyn2013a,SanJose2016a}. In many of these studies complex potentials
were used, which provide an efficient way of introducing gain or loss to the
probability amplitude in quantum mechanics \cite{Moiseyev2011a}. With their
help the time-dependent processes of a decay or growth of a state can be
described by a stationary but non-Hermitian Schr\"odinger equation. Successful
applications to describe a system coupled to an environment in this way can be
found for electromagnetic waves \cite{Klaiman08a,Wiersig2014a,Bittner2014a,%
  Doppler2016a}, electric circuits \cite{Stehmann2004a}, optomechanics
\cite{Xu2016a}, and quantum mechanics \cite{Heiss1999a,Magunov2001a,%
  Latinne1995,Hernandez2006,Graefe10a,Lefebvre2009,Cartarius2011b,Heiss13a,%
  Gao2015a,Menke2016a,Schwarz2015a,Abt2015a}. 

In most studies of edge states in presence of gain or loss, particular
attention was devoted to $\mathcal{PT}$-symmetric potentials, i.e.\
potentials which commute under the combined action of the parity and the
time-reversal operators $\mathcal{P}$ and $\mathcal{T}$, respectively.
$\mathcal{PT}$-symmetric Hamiltonians represent a special class of operators
since they allow for real energy eigenvalues and even completely real energy
spectra even though they are not Hermitian. This is possible since it can be
shown that an eigenvalue of a $\mathcal{PT}$-symmetric operator is always
real if the corresponding eigenstate $\psi$ possesses the same symmetry.
If it does not, there is always a second eigenstate $\mathcal{PT}\psi$ and
the eigenvalues of these two eigenstates form a pair of complex conjugates
\cite{Bender2007a}. The latter case is usually referred to as spontaneously
broken $\mathcal{PT}$ symmetry. In the case of a Hamiltonian the complex energy
eigenvalues describe a growth (positive imaginary part) or decay (negative
imaginary part) of the probability amplitude. The opposite case of preserved
$\mathcal{PT}$ symmetry is then related to the situation of balanced gain and
loss. The in- and outfluxes are spatially separated but have the same strength.
In total the state does neither decay nor grow, and this is expressed by its
real energy eigenvalue. The physical reality of these relations have been proved
experimentally in optics \cite{Rueter10a,Peng2014a,Guo09a}, however, proposals
also exist for Bose-Einstein condensates in quantum mechanics
\cite{Graefe2012a,Single14a,Kreibich14a,Gutoehrlein2015a}.

Together with the Kitaev chain \cite{Kitaev01a} the Su-Schrieffer-Heeger (SSH)
model \cite{Su79a} stands in the main focus when the relation between
$\mathcal{PT}$ symmetry and topologically protected edge states is
investigated. In these models two completely different behaviors were observed.
While in the Kitaev chain it was found that the $\mathcal{PT}$ symmetry is
protected within the TNP when a non-Hermitian potential is applied
\cite{Wang15,Yuce2016a}, the opposite was revealed for the SSH model developed
for the description of the conducting organic material polyacetylene. Its
$\mathcal{PT}$-symmetry is instantaneously broken within the TNP as soon as
gain and loss via a $\mathcal{PT}$-symmetric potential are introduced
\cite{Zhu14a,Yuce2015b}. This can be explained by the different symmetries of
the edge states \cite{Klett2017a}. However, in a modification of the SSH model
new edge states possessing an additional symmetry could be exploited to
experimentally prove the presence of unbroken $\mathcal{PT}$ symmetry in
optics \cite{Weimann2016a}.

The application of complex potentials always means a restriction to an
effective description. The potentials act on the probability amplitude of each
particle to be in the system under consideration \cite{Graefe10a}. A common and
more realistic way of handling environment effects in many-body systems is the
solution of the dynamics using Lindblad master equations \cite{Breuer02a}. With
this description a statistical addition or removal of whole particles is
implemented. It could be shown that this approach is strongly related
to complex $\mathcal{PT}$-symmetric potentials in sufficiently large systems.
For a description of Bose-Einstein condensates with balanced gain and loss of
single particles the mean-field limit is a $\mathcal{PT}$-symmetric
Gross-Pitaevskii equation \cite{Dast14a}. It has been shown that the
topological phases can still be distinguished when information is
extracted from a density matrix resulting from a master equation and that they
can remain robust against certain couplings to the environment
\cite{Viyuela2014a,Linzner2016a}.

In our work we want to address a different question. The purpose of this paper
is to gain insight into the relation of the complex potential introduced above
with the description of particle in- and outcouplings via master equations.
The statistical process introduced in the master equation typically
leads to a crucially different behavior. As soon as the assumption of
smooth gain-loss potentials is violated, in particular in presence of
quantum fluctuations, the stability of $\mathcal{PT}$-symmetric states gets
destroyed as was discussed in optics \cite{Schomerus2010a}. As mentioned
above, for bosonic systems a strong relation between both approaches could
be established. This, however, is possible only in the mean-field limit
\cite{Dast14a}. In the present case we are far away from this limit, and the
temporal evolution of single particles matters. Thus, we cannot expect
that the topological properties of the description with complex potentials
should transfer to the master equation approach. In this paper we find that,
yet, the relation still is very strong.

We investigate this point for the SSH model subject to balanced gain and loss of
particles at different lattice sites. The master equation is formulated in such
a way that it corresponds to the $\mathcal{PT}$-symmetric potentials used in
previous studies \cite{Zhu14a,Klett2017a}, in which it was shown that the
topological edge states break the potential's $\mathcal{PT}$ symmetry.
This should be observable in the dynamics of the master equation via a survival
or decay of an initial occupation of the states. We show that the dynamics
follows the predictions of the stationary calculations accessible in the
effective non-Hermitian approach. In addition, it is possible to see clear
differences in the outcome of the dynamics in cases which correspond to either
the TTP or the TNP in the stationary model. All in all the results indicate that
the observations in the stationary model lead to valuable answers.

We first introduce our model in Sec.\ \ref{sec:model}. Then, in
Sec.\ \ref{sec:dynamics_closed}, we study the dynamics of the closed system
without addition or removal of particles. This provides the basis for
understanding the dynamics in presence of particle gain and loss in Sec.\
\ref{sec:dynamics_open}. The results of the master equation are compared
to the predictions of stationary calculations with complex potentials in
Sec.\ \ref{sec:comparison}. Concluding remarks are given in Sec.\
\ref{sec:conclusion}.

\section{Model and methods}
\label{sec:model}

We consider a one-dimensional case with a lattice distance $a=1$ and a total
of $N$ lattice sites. The system under investigation is the
Su-Schrieffer-Heeger model, of which the Hamiltonian in terms of the fermionic
creation (annihilation) operators $c_i^\dagger$ ($c_i$) reads~\cite{Su79a}
\begin{equation}
  \label{eq:ssh-hamilton}
  H_{\textrm{SSH}} = \sum_{n} \left ( t_- c^\dagger_{2n-1}c_{2n} 
    + t_+ c^\dagger_{2n} c_{2n+1} + \textrm{h.c.} \right )  \; ,
\end{equation}
where $t_{\pm} = t(1\pm \Updelta \cos \Uptheta)$ contains the hopping
amplitude $t$ and the dimerization strength $\Updelta \cos \Uptheta$,
which can vary from $-\Updelta$ to $\Updelta$.

We will investigate the system under environment effects in such a way
that it can
exchange particles (electrons) with the environment at both ends of the chain.
In accordance with \cite{Dast14a} this is done with a master equation for the
density operator $\varrho$ in Lindblad form,
\begin{equation}
  \label{eq:lindblad-masterequation}
  \partial_t \varrho = - \ii [H_{\textrm{SSH}},\varrho] + L_+(\varrho)
  + L_-(\varrho) \; .
\end{equation}
The superoperators read
\begin{eqnarray}
  L_-(\varrho) &=& - \frac{\gamma}{2} (c^\dagger_1 c_1 \varrho
                 + \varrho c^\dagger_1 c_1 - 2 c_1 \varrho c^\dagger_1) \; , 
                 \label{eq:Lm}
\\
  L_+(\varrho) &=& - \frac{\gamma}{2} (c_N c^\dagger_N \varrho 
                 + \varrho c_N c^\dagger_N - 2 c^\dagger_N \varrho c_N) \; ,
                 \label{eq:Lp} 
\end{eqnarray}
where we assume that the possibility to take a particle out of the
system at the first site is the same as the probability of injecting a
particle at the last site of the chain. The strength of the in- and outcoupling
effect is described by $\gamma$.

To solve the Lindblad master equation~(\ref{eq:lindblad-masterequation}) a
quantum Monte-Carlo approach is used \cite{Molmer1996a}, for which the
Hamiltonian has to be represented in matrix form. This limits the size of the
accessible Fock space. However, since the goal of this paper is a comparison
of the results of the Lindblad master equations with those of a non-Hermitian
stationary calculation, which addresses a single-particle problem
\cite{Zhu14a,Yuce2015b,Klett2017a}, we restrict the system to the case that
only a single particle is present on the lattice. To obtain a quantity comparable to the
occupation probabilities of the stationary calculations we calculate the mean
value of the particle number operator at each site,
\begin{equation}
  \label{eq:mean-value}
  \braket{n_i(t)} = \braket{\varrho(t) n_i} ,
\end{equation}
where $i$ is the index of the lattice site. Since the master equation is a
statistical approach and we are interested in a relation to an effective
stationary method  we introduce, in addition,  a temporal mean value defined as
\begin{equation}
  \label{eq:time-mean-value}
  \braket{n_i}_T = \frac{1}{s} \sum_{j=0}^{s} \braket{n_i(t_j)} 
\end{equation}
for the particle number operators $n_i$. The averaging is done over a time
span $T$ discretized in $s$ equal time steps.

\section{Dynamics}
\label{sec:dynamics}

\subsection{Dynamics of the closed system}
\label{sec:dynamics_closed}

To be able to identify signatures of the edge states in the presence of gain
and loss we first investigate their visibility in dynamical calculations
of the closed system. Thus, we neglect the coupling to the environment in this
section, i.e., $\gamma=0$. In this case the topological phase transition point
in the limit of an infinite lattice size $N\to \infty$ can be calculated
analytically using the Zak-phase \cite{Zak1989a}. The transition point between
the TNP and the TTP only depends on the dimerization strength and is $\Uptheta
= \pi/2$. For lower values of $\Uptheta$ the system is in the TNP, for higher
values only topologically trivial states appear.

The Lindblad master equation is a first-order differential equation, and thus
we need an initial state for our calculations. In our study the initial states
are chosen as eigenstates of the time-independent single-particle problem
\cite{Klett2017a}. Since we are mainly interested in the different
topological phases of the SSH model we use the two edge states and one randomly
chosen bulk state as starting points of the temporal evolution. The
initial states are calculated in the TNP since the edge states are only visible
within this parameter regime. Note that this means that the states are not
in all cases stationary states of the system with the given parameters.
This is not necessary for our investigation, since we are only interested in the
differences that appear in the temporal evolution in the two topologically
different regimes.

\begin{figure}[tb]
  \centering
  \includegraphics[width=\columnwidth]{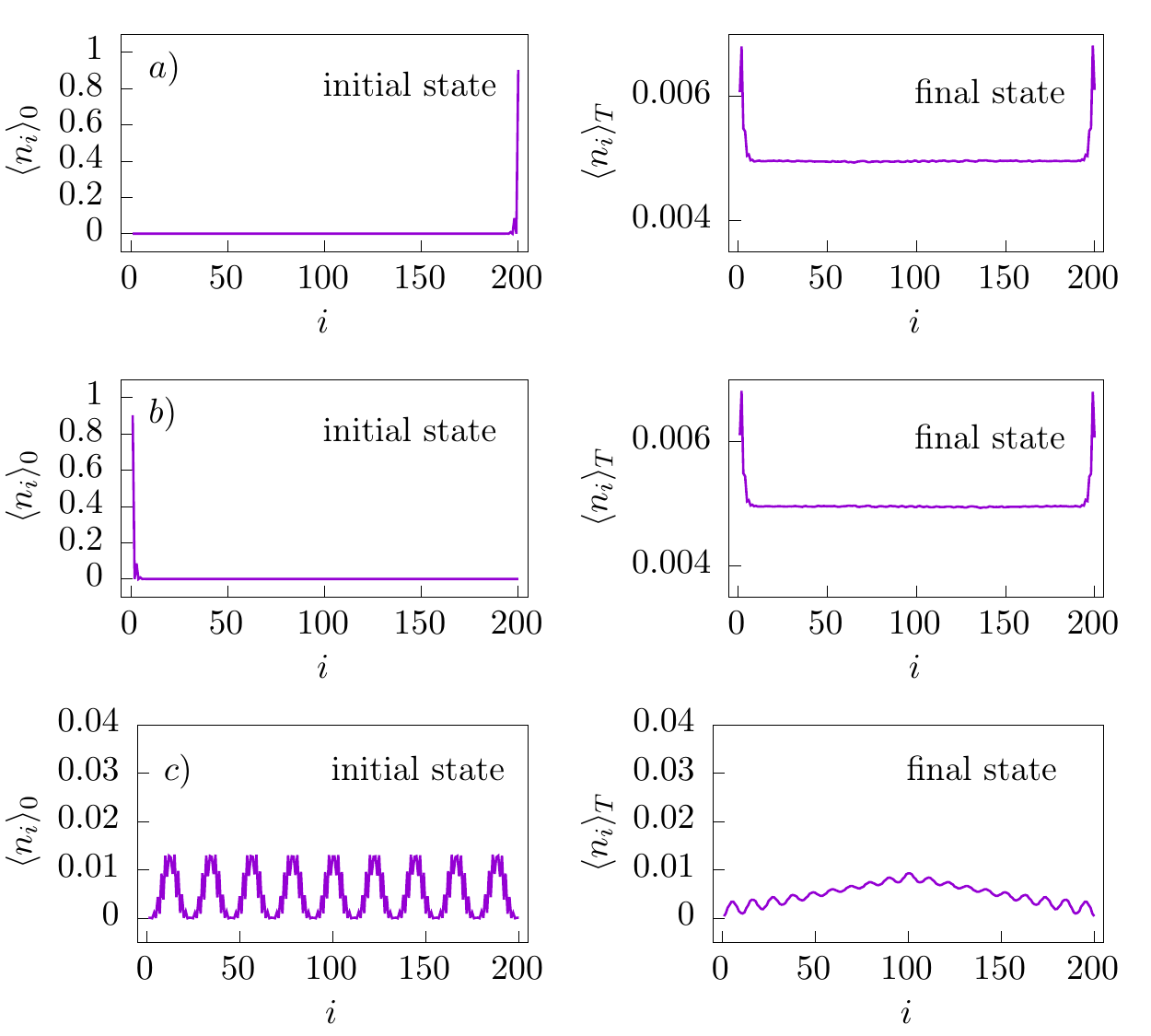}
  \caption{Initial (left column) and corresponding final states (right column)
    for a temporal evolution in the SSH model without external gain and loss
    effects. The dimerization angle has a value of $\Uptheta = 0.9\pi$, and thus
    the system is in the topologically trivial phase, in which edge states
    do not exist. The calculation is carried out with the parameters $N=200$,
    $t=1.0$, and $\Updelta=0.3$, and the time for the evolution is $T=25000$.}
  \label{fig:trivial-time-evolution-no-potential}
\end{figure}

To distinguish the different topological phases of the SSH model we perform
the time evolution of the Lindblad master equation for two different values of
the dimerization strength. Figure \ref{fig:trivial-time-evolution-no-potential}
shows the example for the TTP with the value $\Uptheta = 0.9\pi$. The three
initial states mentioned above are shown in the left column. The final state
corresponding to the initial condition can be found next to it in the right
column. What is shown is the temporal mean value from Eq.\
(\ref{eq:time-mean-value}). In all three cases one can observe that the states
change under the evolution. This is expected since, as mentioned above, they
are not stationary states of the system for this value of $\Uptheta$. The
important observation is that all states obtain a bulk character after the
evolution. In the two cases in which the edge states are used as initial
conditions, a slight predominance of the occupation probability at \emph{both}
edges seems to appear. However, this has to be compared to a broad and almost
uniform distribution of the probability on all lattice sites. In this respect
the excess at the edges is small. The initial bulk distribution involves into a
different one.

\begin{figure}[tb]
  \centering
  \includegraphics[width=\columnwidth]{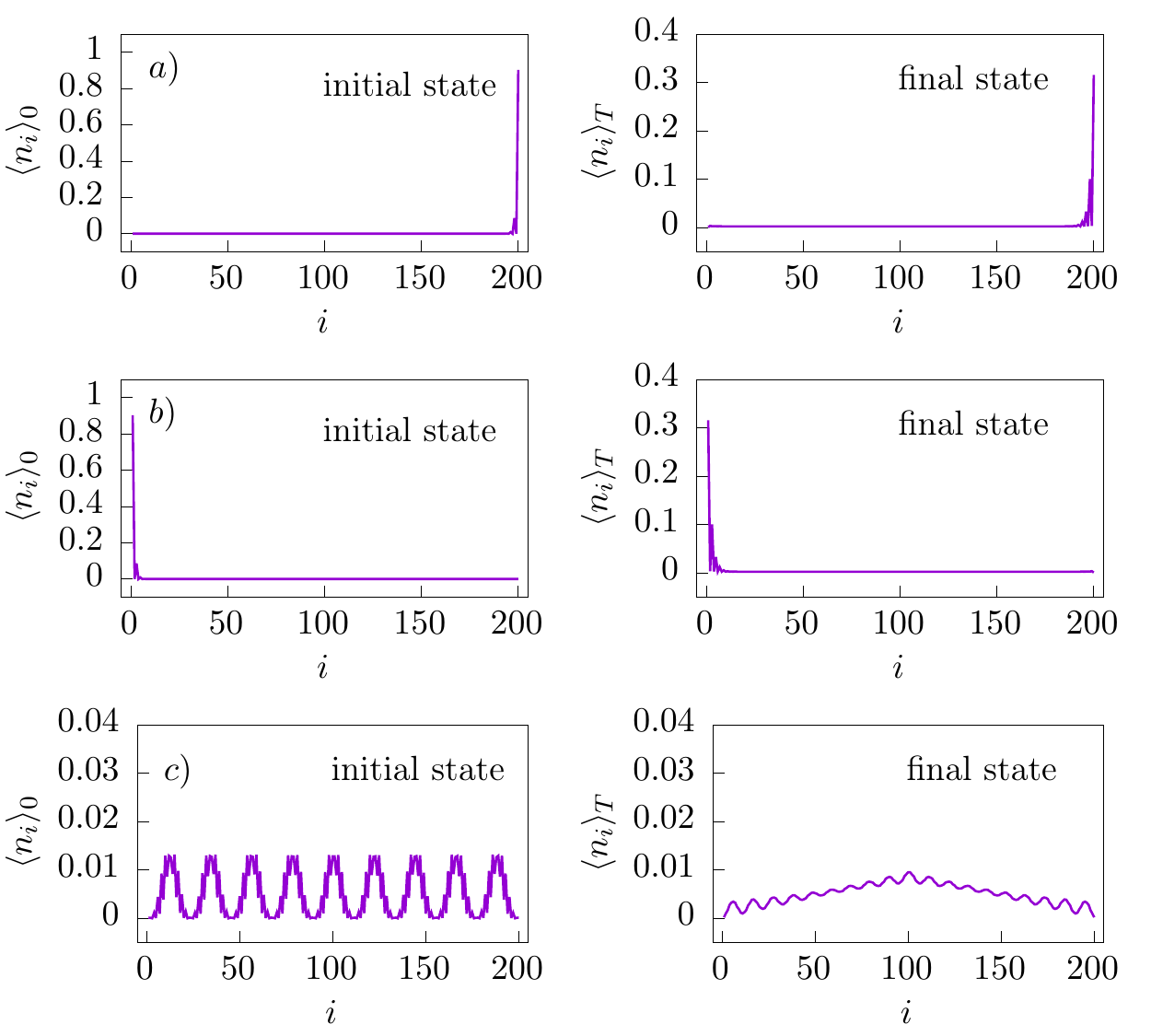}
  \caption{Initial (left column) and final states (right column) for a
    temporal evolution in the SSH model without external gain and loss effects.
    With the dimerization angle $\Uptheta = 0.1\pi$ the system is in the
    topologically nontrivial phase. The remaining parameters are the same as in
    Fig.\ \ref{fig:trivial-time-evolution-no-potential}.}
  \label{fig:nontrivial-time-evolution-no-potential}
\end{figure}

To observe the differences the behavior of the time evolution has to be compared
with that in the TNP, which is shown for $\Uptheta = 0.1\pi$ in Fig.\
\ref{fig:nontrivial-time-evolution-no-potential}.
The initial states are the same as in Fig.\
\ref{fig:trivial-time-evolution-no-potential}. After the time evolution they
show a structure that clearly differs from the previous case. The initial edge
states almost retain their shape. A few lattice sites in the vicinity of the
edges grow in intensity and the edge site itself becomes slightly damped.
However, each of the initial edge states remains clearly located at \emph{one}
edge. The initial bulk state evolves into almost the same state as in the TTP.

\begin{figure}[tb]
  \centering
  \includegraphics[width=\columnwidth]{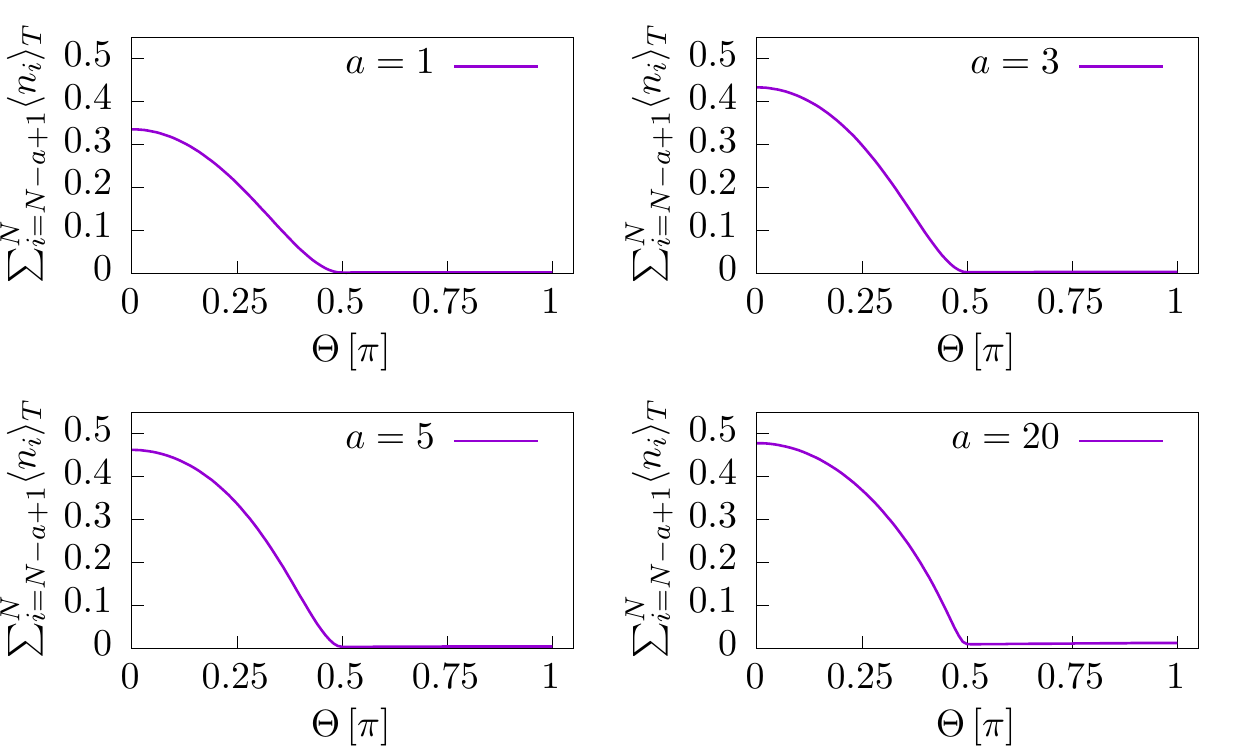}
  \caption{Mean value of the occupation of the last $a$ lattice sites in
    dependence of the dimerization angle $\Uptheta$ without gain and loss. The
    analytical topological phase transition point in the case of $N\to \infty$
    is at $\Uptheta = \pi/2$. One can clearly see that around the analytical
    phase transition point the amplitude of finding a particle at the edge
    is drastically reduced as compared to the TNP. It stays almost constant
    beyond the transition point. The parameters are $\Updelta = 0.3$, $t=1.0$,
    and $N=1000$. The time evolution is done for $T=25000$. In
    all cases the initial state has a distribution mainly localized at the
    right edge, see figure~\ref{fig:trivial-time-evolution-no-potential} (a).}
  \label{fig:phase-diagram-1000}
\end{figure}

The study so far already gives us a hint that it is possible to distinguish the
TTP and the TNP from the dynamics of different initial states. To obtain an
even deeper insight  we plot in Fig.\ \ref{fig:phase-diagram-1000} the occupation
of the last 1, 3, 5, and 20 lattice sites after the temporal evolution as a
function of the dimerization angle $\Uptheta$. 
In all cases the initial state is the edge state localized at the end of the
chain, i.e.\ that in the first row of Fig.\
\ref{fig:trivial-time-evolution-no-potential}.
After the time evolution of $T=25000$ time units the temporal average
(\ref{eq:time-mean-value}) shows the same results in all four cases. For
dimerization angles $\Uptheta \lesssim \pi/2$ we find a nonvanishing
occupation at the edges, which shrinks to almost zero as $\Uptheta$ approaches
$\pi/2$. For larger values of $\Uptheta$ the occupation remains constant at a
very low level. At the critical dimerization strength a sharp kink is visible.
In the case $N\to \infty$ the topological phase transition is exactly at
$\Uptheta = \pi/2$. Thus, the sharp kink is the signature we search for the
identification of the topological phase transition in a dynamical calculation
that is also feasible with in- and outfluxes of particles.

\subsection{Dynamics in presence of particle gain and loss}
\label{sec:dynamics_open}

\begin{figure}[tb]
  \centering
  \includegraphics[width=\columnwidth]{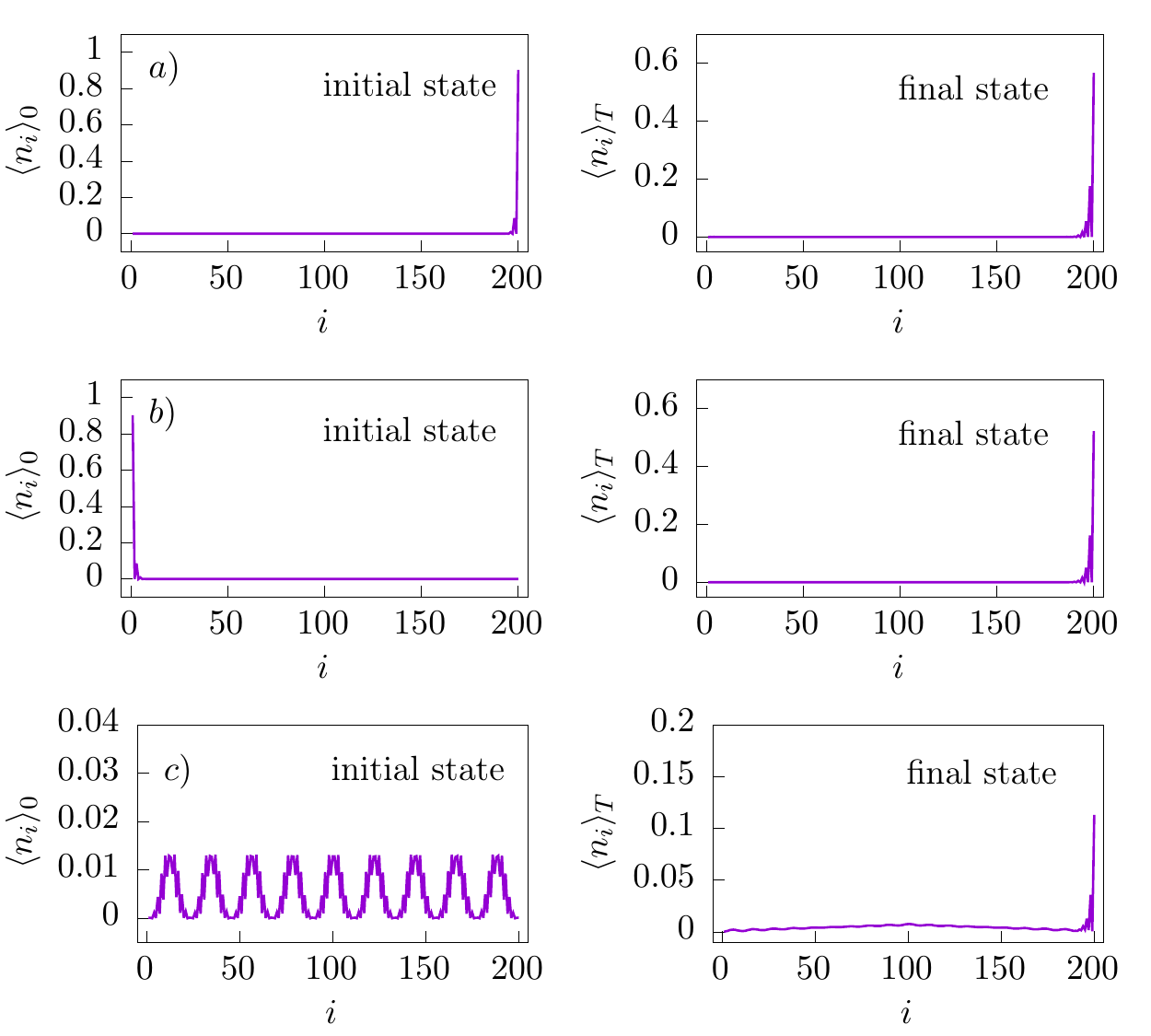}
  \caption{Initial (left column) and final states (right column) as in Fig.\
    \ref{fig:nontrivial-time-evolution-no-potential} with nonvanishing gain
    and loss effect $\gamma = 0.1$. With the dimerization angle
    $\Uptheta = 0.1\pi$ the system is in the topologically nontrivial phase.
    The parameters are $N=200$, $t=1.0$ and $\Updelta=0.3$. The time for
    the evolution is $T=25000$. Dominated by the edge state coupling to the
    influx of particles the particle predominantly occupies the right edge
    for all initial distributions.}
  \label{fig:nontrivial-time-evolution-potential}
\end{figure}

With the method used in the previous section we can turn to the case
of particle gain and loss, i.e.\ we set $\gamma=0.1$ and solve the
Lindblad master equation (\ref{eq:lindblad-masterequation}). An example for the
TNP is shown in Fig.\ \ref{fig:nontrivial-time-evolution-potential}.
As expected for the topologically nontrivial phase edge features are clearly
visible. However, it is always the same edge state which dominates the
evolution independent of the initial state. This is not surprising since
the lattice site with the highest occupation is the last one, i.e. that with
gain (cf.\ Eq.\ (\ref{eq:Lp})). As is known from the stationary calculation
with complex potentials \cite{Zhu14a,Klett2017a} the edge states of the SSH
model break the $\mathcal{PT}$ symmetry of the Hamiltonian, and thus they
cannot exist as stationary states. They will either gain or lose in amplitude,
depending on whether the amplification or the damping lattice site prevails.
Since the initial states chosen in Fig.\
\ref{fig:nontrivial-time-evolution-potential} are no eigenstates of the system
there is always an overlap with the growing state, and thus it will dominate
every temporal evolution. This is what is observed in the figure.

\begin{figure}[tb]
  \centering
  \includegraphics[width=\columnwidth]{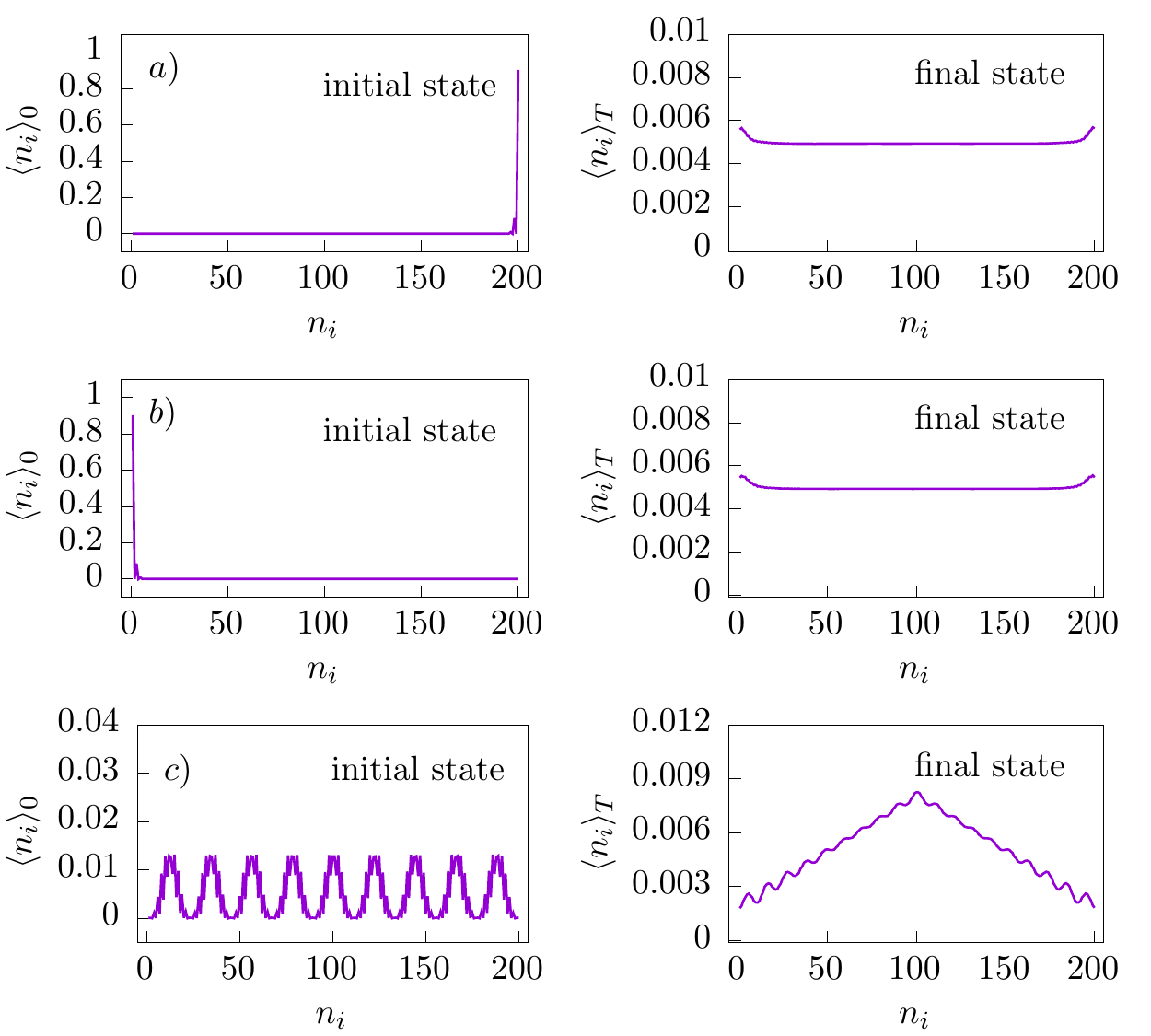}
  \caption{Initial (left column) and final states (right column) in the case
    of gain and loss for $\gamma = 0.1$ and a dimerization angle of $\Uptheta
    = 0.9\pi$, i.e.\ the system is in the topologically trivial phase. The
    remaining parameters are the same as in Fig.\
    \ref{fig:nontrivial-time-evolution-potential}. Since there are no
    $\mathcal{PT}$-broken edge states in the TTP the time averaged occupation
    ends up in a broad and symmetric bulk distribution.}
  \label{fig:trivial-time-evolution-potential}
\end{figure}

Since the amplification is obviously the most significant effect seen in Fig.\
\ref{fig:nontrivial-time-evolution-potential} and only the site with particle
gain stands out one might conclude that the existence of topologically
nontrivial states cannot be identified as soon as the gain effect is present.
If a site with gain exists and is occupied (or disproportionately
highly occupied as compared to loss sites due to an asymmetry) it can be
expected to dominate the whole dynamics. This should give rise to a large
occupation at the edge that is not distinguishable from an edge state.
But this assumption turns out to be wrong. The TTP with the same gain and
loss strength is shown in Fig.\ \ref{fig:trivial-time-evolution-potential},
in which a clear difference to Fig.\
\ref{fig:nontrivial-time-evolution-potential} is visible. Now the dynamics ends
up in a bulk state for all initial states. For both edge states the same final
distribution is obtained. There are still slightly higher occupations at the
edges than in the center of the bulk, but they are drastically weaker as
compared to the Hermitian case (cf.\ Fig.\
\ref{fig:trivial-time-evolution-no-potential}).

The missing dominance of the edge states can be explained in a simple way.
From the stationary calculation we know that they are the first states breaking
the $\mathcal{PT}$ symmetry, and for the chosen $\gamma = 0.1$ they are even
the only states for which gain and loss are not balanced. Consequently, in
the TTP, in which no edge states are present, the whole spectrum consists of
states experiencing a unitary time evolution. Any initial state can be
decomposed into these stationary states and will undergo oscillations with
constant total amplitude. The time average of these oscillations
assumes the symmetric distributions visible in Fig.\
\ref{fig:trivial-time-evolution-potential}. Thus, it is the strong relation of
the edge states to $\mathcal{PT}$-symmetry breaking in the SSH model, which
introduces a pronounced difference in the dynamics.

\begin{figure}[tb]
  \centering
  \includegraphics[width=\columnwidth]{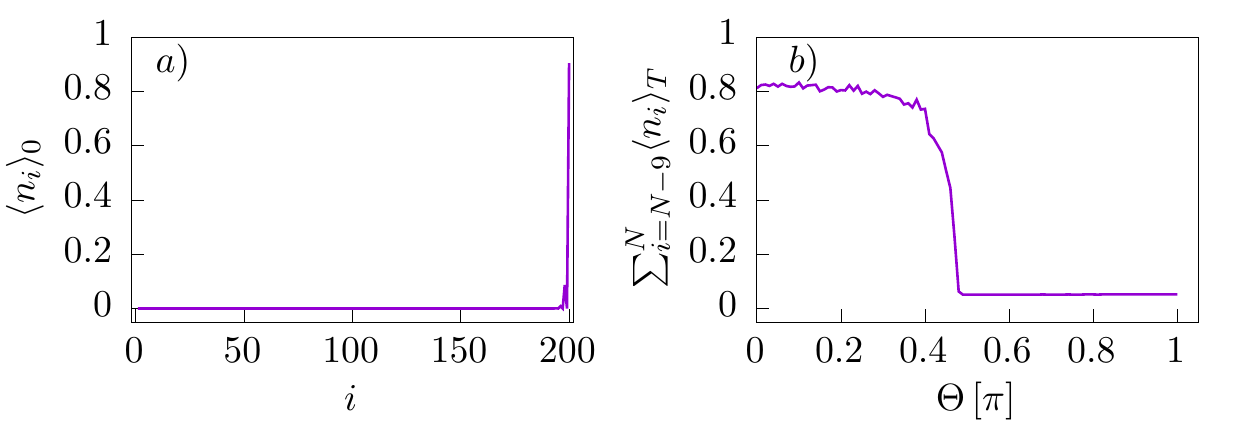}
  \caption{Mean value b) of the occupation of the last 10 lattice sites in
    dependence of the dimerization angle $\Uptheta$ in presence of gain and loss
    with the value $\gamma = 0.1$. The initial state is shown in a). The
    remaining parameters are $\Updelta = 0.3$, $t=1.0$, and $N=1000$. The time
    evolution is done for $T=25000$. In comparison to the case without gain
    and loss the kink in the edge occupation is much more pronounced, cf.\
    Fig.\ \ref{fig:phase-diagram-1000}.}
    \label{fig:phase-diagram-potential}
\end{figure}

Since the presence of the edge states has a strong influence on the dynamics,
they also have a significant impact on the time averaged occupation at the
right edge as can be seen in Fig.\ \ref{fig:phase-diagram-potential}.
For the initial state shown on the left, the time averaged mean occupation of
the last 10 lattice sites is drawn on the right as a function of $\Uptheta$.
The result is similar to that in Fig.\ \ref{fig:phase-diagram-1000}. However,
the kink at the topological phase transition is much sharper. Thus, it still
clearly identifies the topological phase transition. It appears just below
$\Uptheta = \pi/2$, i.e.\ at the point, where it also appears in the case without
in- and outcoupling of particles.

\subsection{Comparison to the stationary non-Hermitian approach}
\label{sec:comparison}

An important question in this study is the comparability of the results of
the non-hermitian stationary
calculation using complex potentials \cite{Zhu14a,Klett2017a} with the
more realistic in- and outcoupling of particles described by Lindblad master
equations. The procedure used in this paper was capable of identifying the
appearance of edge states, and thus of distinguishing the two topological
phases. From the stationary calculations it is known that topologically nontrivial
edge states can be found even in the presence of gain and loss effects, and the
phase transition point is not affected by gain and loss acting on the edge
sites \cite{Zhu14a,Klett2017a}. The same is now found in the dynamical
calculations with master equations. The difference to the analytical value
$\Uptheta = \pi/2$ can be traced back to the finite lattice size.

The agreement between both approaches goes even beyond the pure existence of
the edge states for the same values of the dimerization angle $\Uptheta$. The
dynamics of the master equations follows exactly the predictions that can be
done by looking at the existence or nonexistence of complex energy eigenstates
in the stationary picture. This turned out to be valuable in the explanation
of the final states in Sec.\ \ref{sec:dynamics_open}. The influence of the
complex energy eigenstates is clearly present.

\section{Conclusion}
\label{sec:conclusion}

We studied the SSH model with gain and loss at the edges, which was introduced
via Lindblad master equations. This description is more realistic than that of
previous studies using effective complex $\mathcal{PT}$-symmetric potentials
\cite{Zhu14a,Klett2017a}. We found that the addition and removal of particles
does not destroy the existence of two distinct topological phases identified
by the existence of edge states. These edge states could be detected in the
dynamics by studying time-averaged occupation probabilities of the individual
lattice sites. This is in very good agreement with calculations of the
stationary Schr\"odinger equation in presence of complex potentials.

The agreement goes even further. The dynamical predictions of the energy
eigenvalues of the stationary approach, which are real in the case of
$\mathcal{PT}$-symmetric states and complex for broken $\mathcal{PT}$ symmetry,
can be clearly observed in the evolution of the density operator using the
master equations. The study shows that both approaches agree very well. In the
investigation of Bose-Einstein condensates it was shown that a Gross-Pitaevskii
equation with complex potentials is the exact mean-field limit of a
many-particle description, in which gain and loss are implemented via Lindblad
master equations \cite{Dast14a,Dast2017a}. In this study we have found that even
if both approaches act on the individual particles, a good agreement is found,
and thus the application of complex potentials in many works dealing with
topological condensed matter models subject to gain and loss
\cite{Hu2011a,Esaki2011a,Ghosh2012a,Schomerus2013a,Zeuner2015a,Yuce2015a,%
  Zhu14a,Yuce2015b,Wang15,Yuce2016a} is justified.

The identification of the two different topological phases is already very
clear by investigating the dynamics. However, one certainly is interested
in finding a more direct method by directly determining a topological
invariant \cite{Linzner2016a,Bardyn2013a,Lieu2018a,Wagner2017a}. Thus,
a natural next step is the extension of the formalism in such a way that a
topological invariant, e.g.\ an extended complex Zak phase, can be obtained.
In addition, the current study was restricted to the single-particle case. In
presence of gain the particle number can in principle change. Thus, it would
be interesting to see how the system behaves if more particles can enter. This
would be of crucial importance for bosons, which can occupy the site with gain
in large numbers, and thus easily introduce instabilities in the dynamics.

\section*{Acknowledgements}
  We acknowledge financial support from the Deutsche For\-schungs\-ge\-mein\-schaft
  (DFG) through ZUK 63.

  \section*{Author contribution statement}
  All authors contributed equally to this work
  

\begin{thebibliography}{10}

\bibitem{Hasan2010a}
M.~Z. Hasan and C.~L. Kane.
\newblock \textit{Colloquium}: Topological insulators.
\newblock {\em Rev. Mod. Phys.}, 82:3045--3067, 2010.

\bibitem{Alicea12a}
Jason Alicea.
\newblock New directions in the pursuit of {M}ajorana fermions in solid state
  systems.
\newblock {\em Rep. Prog. Phys.}, 75(7):076501, 2012.

\bibitem{Altland1997a}
Alexander Altland and Martin~R. Zirnbauer.
\newblock Nonstandard symmetry classes in mesoscopic normal-superconducting
  hybrid structures.
\newblock {\em Phys. Rev. B}, 55:1142--1161, 1997.

\bibitem{Schnyder2008a}
Andreas~P. Schnyder, Shinsei Ryu, Akira Furusaki, and Andreas W.~W. Ludwig.
\newblock Classification of topological insulators and superconductors in three
  spatial dimensions.
\newblock {\em Phys. Rev. B}, 78:195125, 2008.

\bibitem{Leijnse2012a}
Martin Leijnse and Karsten Flensberg.
\newblock Introduction to topological superconductivity and {M}ajorana
  fermions.
\newblock {\em Semicond. Sci. Technol.}, 27(12):124003, 2012.

\bibitem{Klitzing1980a}
K.~v. Klitzing, G.~Dorda, and M.~Pepper.
\newblock New method for high-accuracy determination of the fine-structure
  constant based on quantized {Ha}ll resistance.
\newblock {\em Phys. Rev. Lett.}, 45:494--497, 1980.

\bibitem{Klitzing1986a}
Klaus von Klitzing.
\newblock The quantized {Ha}ll effect.
\newblock {\em Rev. Mod. Phys.}, 58:519--531, 1986.

\bibitem{Thouless1982a}
D.~J. Thouless, M.~Kohmoto, M.~P. Nightingale, and M.~den Nijs.
\newblock Quantized {Ha}ll conductance in a two-dimensional periodic potential.
\newblock {\em Phys. Rev. Lett.}, 49:405--408, 1982.

\bibitem{Mourik2012a}
V.~Mourik, K.~Zuo, S.~M. Frolov, S.~R. Plissard, E.~P. A.~M. Bakkers, and L.~P.
  Kouwenhoven.
\newblock Signatures of {M}ajorana fermions in hybrid
  superconductor-semiconductor nanowire devices.
\newblock {\em Science}, 336:1003--1007, 2012.

\bibitem{Stanescu2013a}
T.~D. Stanescu and S.~Tewari.
\newblock {M}ajorana fermions in semiconductor nanowires: fundamentals,
  modeling, and experiment.
\newblock {\em J. Phys.: Condens. Matter}, 25(23):233201, 2013.

\bibitem{Elliott2015a}
Steven~R. Elliott and Marcel Franz.
\newblock Colloquium: Majorana fermions in nuclear, particle, and solid-state
  physics.
\newblock {\em Rev. Mod. Phys.}, 87:137--163, 2015.

\bibitem{Carmele2015a}
Alexander Carmele, Markus Heyl, Christina Kraus, and Marcello Dalmonte.
\newblock Stretched exponential decay of {M}ajorana edge modes in many-body
  localized {K}itaev chains under dissipation.
\newblock {\em Phys. Rev. B}, 92:195107, 2015.

\bibitem{Hu2011a}
Yi~Chen Hu and Taylor~L. Hughes.
\newblock Absence of topological insulator phases in non-hermitian
  $\mathcal{PT}$-symmetric hamiltonians.
\newblock {\em Phys. Rev. B}, 84:153101, 2011.

\bibitem{Esaki2011a}
Kenta Esaki, Masatoshi Sato, Kazuki Hasebe, and Mahito Kohmoto.
\newblock Edge states and topological phases in non-hermitian systems.
\newblock {\em Phys. Rev. B}, 84:205128, 2011.

\bibitem{Ghosh2012a}
Pijush~K. Ghosh.
\newblock A note on the topological insulator phase in non-hermitian quantum
  systems.
\newblock {\em J. Phys.: Condens. Matter}, 24(14):145302, 2012.

\bibitem{Viyuela2012a}
O.~Viyuela, A.~Rivas, and M.~A. Martin-Delgado.
\newblock Thermal instability of protected end states in a one-dimensional
  topological insulator.
\newblock {\em Phys. Rev. B}, 86:155140, 2012.

\bibitem{Schomerus2013a}
Henning Schomerus.
\newblock Topologically protected midgap states in complex photonic lattices.
\newblock {\em Opt. Lett.}, 38(11):1912--1914, 2013.

\bibitem{Rivas2013a}
A.~Rivas, O.~Viyuela, and M.~A. Martin-Delgado.
\newblock Density-matrix {C}hern insulators: Finite-temperature generalization
  of topological insulators.
\newblock {\em Phys. Rev. B}, 88:155141, 2013.

\bibitem{Zeuner2015a}
Julia~M. Zeuner, Mikael~C. Rechtsman, Yonatan Plotnik, Yaakov Lumer, Stefan
  Nolte, Mark~S. Rudner, Mordechai Segev, and Alexander Szameit.
\newblock Observation of a topological transition in the bulk of a
  non-hermitian system.
\newblock {\em Phys. Rev. Lett.}, 115:040402, 2015.

\bibitem{Yuce2015a}
C.~Yuce.
\newblock Topological phase in a non-hermitian symmetric system.
\newblock {\em Phys. Lett. A}, 379(18–19):1213 -- 1218, 2015.

\bibitem{Zhu14a}
Baogang Zhu, Rong L\"u, and Shu Chen.
\newblock $\mathcal{PT}$ symmetry in the non-hermitian
  {S}u-{S}chrieffer-{H}eeger model with complex boundary potentials.
\newblock {\em Phys. Rev. A}, 89:062102, 2014.

\bibitem{Yuce2015b}
Cem Yuce.
\newblock $\mathcal{PT}$ symmetric {F}loquet topological phase.
\newblock {\em Eur. Phys. J. D}, 69(7):184, 2015.

\bibitem{Wang15}
Xiaohui Wang, Tingting Liu, Ye~Xiong, and Peiqing Tong.
\newblock Spontaneous $\mathcal{PT}$-symmetry breaking in non-hermitian
  {K}itaev and extended {K}itaev models.
\newblock {\em Phys. Rev. A}, 92:012116, 2015.

\bibitem{Yuce2016a}
C.~Yuce.
\newblock {M}ajorana edge modes with gain and loss.
\newblock {\em Phys. Rev. A}, 93:062130, 2016.

\bibitem{Benito2016a}
M\'onica Benito, Michael Niklas, Gloria Platero, and Sigmund Kohler.
\newblock Edge-state blockade of transport in quantum dot arrays.
\newblock {\em Phys. Rev. B}, 93:115432, 2016.

\bibitem{Niklas2016a}
Michael Niklas, Mónica Benito, Sigmund Kohler, and Gloria Platero.
\newblock Transport, shot noise, and topology in ac-driven dimer arrays.
\newblock {\em Nanotechnology}, 27(45):454002, 2016.

\bibitem{Sedlmayr2018a}
N.~Sedlmayr, M.~Fleischhauer, and J.~Sirker.
\newblock Fate of dynamical phase transitions at finite temperatures and in
  open systems.
\newblock {\em Phys. Rev. B}, 97:045147, 2018.

\bibitem{Sedlmayr2018b}
N.~Sedlmayr, P.~Jaeger, M.~Maiti, and J.~Sirker.
\newblock Bulk-boundary correspondence for dynamical phase transitions in
  one-dimensional topological insulators and superconductors.
\newblock {\em Phys. Rev. B}, 97:064304, 2018.

\bibitem{Bardyn2013a}
C.-E. Bardyn, M.~A. Baranov, C.~V. Kraus, E.~Rico, A.~Imamoglu, P.~Zoller, and
  S.~Diehl.
\newblock Topology by dissipation.
\newblock {\em New J. Phys.}, 15(8):085001, 2013.

\bibitem{SanJose2016a}
Pablo San-Jose, Jorge Cayao, Elsa Prada, and Ram{\'o}n Aguado.
\newblock {{M}ajorana bound states from exceptional points in non-topological
  superconductors}.
\newblock {\em Sci. Rep.}, 6:21427, 2016.

\bibitem{Moiseyev2011a}
Nimrod Moiseyev.
\newblock {\em Non-Hermitian Quantum Mechanics}.
\newblock Cambridge University Press, Cambridge, 2011.

\bibitem{Klaiman08a}
Shachar Klaiman, Uwe G{\"{u}}nther, and Nimrod Moiseyev.
\newblock Visualization of branch points in $\mathcal{PT}$-symmetric
  waveguides.
\newblock {\em Phys. Rev. Lett.}, 101:080402, 2008.

\bibitem{Wiersig2014a}
Jan Wiersig.
\newblock Enhancing the sensitivity of frequency and energy splitting detection
  by using exceptional points: Application to microcavity sensors for
  single-particle detection.
\newblock {\em Phys. Rev. Lett.}, 112:203901, 2014.

\bibitem{Bittner2014a}
S.~Bittner, B.~Dietz, H.~L. Harney, M.~Miski-Oglu, A.~Richter, and
  F.~Sch\"afer.
\newblock Scattering experiments with microwave billiards at an exceptional
  point under broken time-reversal invariance.
\newblock {\em Phys. Rev. E}, 89:032909, 2014.

\bibitem{Doppler2016a}
J{\"o}rg Doppler, Alexei~A. Mailybaev, Julian B{\"o}hm, Ulrich Kuhl, Adrian
  Girschik, Florian Libisch, Thomas~J. Milburn, Peter Rabl, Nimrod Moiseyev,
  and Stefan Rotter.
\newblock Dynamically encircling an exceptional point for asymmetric mode
  switching.
\newblock {\em Nature}, 537(7618):76--79, 2016.

\bibitem{Stehmann2004a}
T.~Stehmann, W.~D. Heiss, and F.~G. Scholtz.
\newblock Observation of exceptional points in electronic circuits.
\newblock {\em J. Phys. A}, 37(31):7813, 2004.

\bibitem{Xu2016a}
H.~Xu, D.~Mason, Luyao Jiang, and J.~G.~E. Harris.
\newblock Topological energy transfer in an optomechanical system with
  exceptional points.
\newblock {\em Nature}, 537(7618):80--83, 2016.

\bibitem{Heiss1999a}
W.~D. Heiss.
\newblock Phases of wave functions and level repulsion.
\newblock {\em Eur. Phys. J. D}, 7(1):1, 1999.

\bibitem{Magunov2001a}
A.~I. Magunov, I.~Rotter, and S.~I. Strakhova.
\newblock Laser-induced continuum structures and double poles of the
  {S}-matrix.
\newblock {\em J. Phys. B}, 34(1):29, 2001.

\bibitem{Latinne1995}
O.~Latinne, N.~J. Kylstra, M.~D\"orr, J.~Purvis, M.~Terao-Dunseath, C.~J.
  Joachain, P.~G. Burke, and C.~J. Noble.
\newblock Laser-induced degeneracies involving autoionizing states in complex
  atoms.
\newblock {\em Phys. Rev. Lett.}, 74(1):46, 1995.

\bibitem{Hernandez2006}
E~Hern{\'{a}}ndez, A~J{\'{a}}uregui, and A~Mondrag{\'{a}}n.
\newblock Non-hermitian degeneracy of two unbound states.
\newblock {\em J. Phys. A}, 39(32):10087, 2006.

\bibitem{Graefe10a}
Eva-Maria Graefe, Hans~J\"urgen Korsch, and Astrid~Elisa Niederle.
\newblock Quantum-classical correspondence for a non-hermitian {B}ose-{H}ubbard
  dimer.
\newblock {\em Phys. Rev. A}, 82:013629, 2010.

\bibitem{Lefebvre2009}
R.~Lefebvre, O.~Atabek, M.~{\v S}indelka, and N.~Moiseyev.
\newblock Resonance coalescence in molecular photodissociation.
\newblock {\em Phys. Rev. Lett.}, 103(12):123003, 2009.

\bibitem{Cartarius2011b}
Holger Cartarius and Nimrod Moiseyev.
\newblock Fingerprints of exceptional points in the survival probability of
  resonances in atomic spectra.
\newblock {\em Phys. Rev. A}, 84(1):013419, 2011.

\bibitem{Heiss13a}
W.~D. Heiss, H.~Cartarius, G.~Wunner, and J.~Main.
\newblock Spectral singularities in $\mathcal{PT}$-symmetric {B}ose-{E}instein
  condensates.
\newblock {\em J. Phys. A}, 46(27):275307, 2013.

\bibitem{Gao2015a}
T.~Gao, E.~Estrecho, K.~Y. Bliokh, T.~C.~H. Liew, M.~D. Fraser, S.~Brodbeck,
  M.~Kamp, C.~Schneider, S.~Hofling, Y.~Yamamoto, F.~Nori, Y.~S. Kivshar, A.~G.
  Truscott, R.~G. Dall, and E.~A. Ostrovskaya.
\newblock Observation of non-hermitian degeneracies in a chaotic
  exciton-polariton billiard.
\newblock {\em Nature}, 526(7574):554--558, 2015.

\bibitem{Menke2016a}
Henri Menke, Marcel Klett, Holger Cartarius, J\"org Main, and G\"unter Wunner.
\newblock State flip at exceptional points in atomic spectra.
\newblock {\em Phys. Rev. A}, 93:013401, 2016.

\bibitem{Schwarz2015a}
Lukas Schwarz, Holger Cartarius, G{\"u}nter Wunner, Walter~Dieter Heiss, and
  J{\"o}rg Main.
\newblock Fano resonances in scattering: An alternative perspective.
\newblock {\em Eur. Phys. J. D}, 69(8), 2015.

\bibitem{Abt2015a}
Nikolas Abt, Holger Cartarius, and G{\"u}nter Wunner.
\newblock Supersymmetric model of a {B}ose-{E}instein condensate in a
  $\mathcal{PT}$-symmetric double-delta trap.
\newblock {\em Int. J. Theor. Phys.}, 54(11):4054--4067, 2015.

\bibitem{Bender2007a}
Carl~M. Bender.
\newblock Making sense of non-hermitian hamiltonians.
\newblock {\em Rep. Progr. Phys.}, 70(6):947, 2007.

\bibitem{Rueter10a}
Christian~E. R{\"{u}}ter, Konstantinos~G. Makris, Ramy El-Ganainy, Demetrios~N.
  Christodoulides, Mordechai Segev, and Detlev Kip.
\newblock Observation of parity-time symmetry in optics.
\newblock {\em Nat. Phys.}, 6:192, 2010.

\bibitem{Peng2014a}
Bo~Peng, Sahin~Kaya Ozdemir, Fuchuan Lei, Faraz Monifi, Mariagiovanna
  Gianfreda, Gui~Lu Long, Shanhui Fan, Franco Nori, Carl~M. Bender, and Lan
  Yang.
\newblock Parity-time-symmetric whispering-gallery microcavities.
\newblock {\em Nat. Phys.}, 10(5):394--398, 2014.

\bibitem{Guo09a}
A.~Guo, G.~J. Salamo, D.~Duchesne, R.~Morandotti, M.~Volatier-Ravat, V.~Aimez,
  G.~A. Siviloglou, and D.~N. Christodoulides.
\newblock Observation of $\mathcal{PT}$-symmetry breaking in complex optical
  potentials.
\newblock {\em Phys. Rev. Lett.}, 103:093902, 2009.

\bibitem{Graefe2012a}
Eva-Maria Graefe.
\newblock Stationary states of a $\mathcal{PT}$ symmetric two-mode
  {B}ose-{E}instein condensate.
\newblock {\em J. Phys. A}, 45(44):444015, 2012.

\bibitem{Single14a}
Fabian Single, Holger Cartarius, G\"unter Wunner, and J\"org Main.
\newblock Coupling approach for the realization of a $\mathcal{PT}$-symmetric
  potential for a {B}ose-{E}instein condensate in a double well.
\newblock {\em Phys. Rev. A}, 90:042123, 2014.

\bibitem{Kreibich14a}
Manuel Kreibich, J\"org Main, Holger Cartarius, and G\"unter Wunner.
\newblock Realizing $\mathcal{PT}$-symmetric non-hermiticity with ultracold
  atoms and hermitian multiwell potentials.
\newblock {\em Phys. Rev. A}, 90:033630, 2014.

\bibitem{Gutoehrlein2015a}
Robin Gut{\"o}hrlein, Jan Schnabel, Ibrokhim Iskandarov, Holger Cartarius,
  J{\"o}rg Main, and G{\"u}nter Wunner.
\newblock Realizing $\mathcal{PT}$-symmetric {BEC} subsystems in closed
  hermitian systems.
\newblock {\em J. Phys. A}, 48(33):335302, 2015.

\bibitem{Kitaev01a}
A.~Yu. Kitaev.
\newblock Unpaired {M}ajorana fermions in quantum wires.
\newblock {\em Sov. Phys.-Usp.}, 44:131, 2001.

\bibitem{Su79a}
W.~P. Su, J.~R. Schrieffer, and A.~J. Heeger.
\newblock Solitons in polyacetylene.
\newblock {\em Phys. Rev. Lett.}, 42:1698--1701, 1979.

\bibitem{Klett2017a}
Marcel Klett, Holger Cartarius, Dennis Dast, J{\"o}rg Main, and G{\"u}nter
  Wunner.
\newblock Relation between $\mathcal{PT}$-symmetry breaking and topologically
  nontrivial phases in the {S}u-{S}chrieffer-{H}eeger and {K}itaev models.
\newblock {\em Phys. Rev. A}, 95:053626, 2017.

\bibitem{Weimann2016a}
S.~Weimann, M.~Kremer, Y.~Plotnik, Y.~Lumer, S.~Nolte, K.~G. Makris, M.~Segev,
  M.~C. Rechtsman, and A.~Szameit.
\newblock {Topologically protected bound states in photonic
  parity-time-symmetric crystals}.
\newblock {\em Nat. Mater.}, 16(4):433--438, 2017.

\bibitem{Breuer02a}
Heinz-Peter Breuer and Francesco Petruccione.
\newblock {\em The Theory of Open Quantum Systems}.
\newblock Oxford University Press, Oxford, 2002.

\bibitem{Dast14a}
Dennis Dast, Daniel Haag, Holger Cartarius, and G\"unter Wunner.
\newblock Quantum master equation with balanced gain and loss.
\newblock {\em Phys. Rev. A}, 90:052120, 2014.

\bibitem{Viyuela2014a}
O.~Viyuela, A.~Rivas, and M.~A. Martin-Delgado.
\newblock Uhlmann phase as a topological measure for one-dimensional fermion
  systems.
\newblock {\em Phys. Rev. Lett.}, 112:130401, 2014.

\bibitem{Linzner2016a}
D.~Linzner, L.~Wawer, F.~Grusdt, and M.~Fleischhauer.
\newblock Reservoir-induced {T}houless pumping and symmetry-protected
  topological order in open quantum chains.
\newblock {\em Phys. Rev. B}, 94:201105, 2016.

\bibitem{Schomerus2010a}
Henning Schomerus.
\newblock Quantum noise and self-sustained radiation of
  $\mathcal{P}\mathcal{T}$-symmetric systems.
\newblock {\em Phys. Rev. Lett.}, 104:233601, 2010.

\bibitem{Molmer1996a}
Klaus M{\o}lmer, Yvan Castin, and Jean Dalibard.
\newblock Monte {C}arlo wave-function method in quantum optics.
\newblock {\em J. Opt. Soc. Am. B}, 10(3):524--538, 1993.

\bibitem{Zak1989a}
J.~Zak.
\newblock Berry's phase for energy bands in solids.
\newblock {\em Phys. Rev. Lett.}, 62:2747--2750, 1989.

\bibitem{Dast2017a}
Dennis Dast, Daniel Haag, Holger Cartarius, J{\"o}rg Main, and G{\"u}nter
  Wunner.
\newblock Stationary states in the many-particle description of
  {B}ose-{E}instein condensates with balanced gain and loss.
\newblock {\em Phys. Rev. A}, 96:023625, 2017.

\bibitem{Lieu2018a}
Simon Lieu.
\newblock Topological phases in the non-hermitian {S}u-{S}chrieffer-{H}eeger
  model.
\newblock {\em Phys. Rev. B}, 97:045106, 2018.

\bibitem{Wagner2017a}
Marcel Wagner, Felix Dangel, Holger Cartarius, G{\"u}nter Wunner, and J{\"o}rg
  Main.
\newblock Numerical calculation of the complex {B}erry phase in non-hermitian
  systems.
\newblock {\em Acta Polytech.}, 57(6):470--476, 2017.

\end{thebibliography}

\end{document}